\pdfoutput=1
\documentclass[twoside,twocolumn,9pt,a4paper]{article}
\usepackage{cmap}
\usepackage{extsizes}
\usepackage[super,sort&compress,comma]{natbib}
\usepackage[version=3]{mhchem}
\usepackage[left=1.65cm, right=1.65cm, top=1.8cm, bottom=2.0cm]{geometry}
\usepackage{balance}
\usepackage{widetext}
\usepackage{sectsty}
\usepackage{graphicx}
\usepackage{lastpage}
\usepackage[format=plain,justification=raggedright,singlelinecheck=false,font={stretch=1.125,small},labelfont=bf,labelsep=space]{caption}

\usepackage{float}
\usepackage{fancyhdr}
\usepackage{fnpos}
\usepackage[british]{babel}
\usepackage{array}
\usepackage{droidsans}
\usepackage{charter}
\usepackage[T1]{fontenc}
\usepackage[usenames,dvipsnames]{xcolor}
\usepackage{setspace}
\usepackage[compact]{titlesec}

\definecolor{cream}{RGB}{222,217,201}

\usepackage{mathptmx}
\usepackage{times}
\usepackage{fixmath}
\usepackage[Euler]{upgreek}
\DeclareSymbolFont{UPM}{U}{eur}{m}{n}  
\DeclareMathSymbol{\partial}{0}{UPM}{"40}
\usepackage[italic]{mathastext}
\usepackage{helvet}

\usepackage[pdftex, bookmarks, pdffitwindow=false, pdfstartview=FitH, pdfdisplaydoctitle, colorlinks, plainpages=false,pdfauthor={Aleks Reinhardt and Daan Frenkel}, pdftitle={DNA brick self-assembly with an off-lattice potential}, pdfpagelabels, hypertexnames, citecolor={blue},linkcolor={red}, urlcolor={violet}, pdflang={en}, hyperfootnotes=false, breaklinks]{hyperref}

\newcommand{\der}{\ensuremath{\mathrm{d}}}
\newcommand{\e}{\ensuremath{\mathrm{e}}} 
\def\abs#1{\ensuremath{\left| #1 \right|}}
\makeatletter
\providecommand*{\standardstate}{{\ensuremath{\protect\cst@sstate}}}
\newcommand*{\cst@sstate}{\mathpalette\cst@s@state\circ}
\newcommand*{\cst@s@state}[2]{\ooalign{\hfil$#1-$\hfil\cr\hfil$#1#2$\hfil\cr}}
\newcommand{\std}{\ensuremath{^\standardstate}}
\makeatother
\newcommand{\pd}[2]{\mathchoice
  {\ensuremath{ \frac{\partial #1}{\partial #2}}}
  {\ensuremath{\partial #1 / \partial #2}}
  {\ensuremath{\partial #1 / \partial #2}}
  {\ensuremath{\partial #1 / \partial #2}}
}

\usepackage{mciteplus}
\usepackage{mleftright}

\usepackage{booktabs}

\usepackage{flafter}
\usepackage{graphicx}
\usepackage{amsmath}

\usepackage[dvipsnames]{xcolor}

\usepackage[stretch=15,shrink=15,step=1]{microtype}
\usepackage{siunitx}
\newcommand{\refSub}[2]{\hyperref[#2]{\ref{#2}(#1)}}

\begin{document}

\pagestyle{fancy}
\thispagestyle{plain}
\fancypagestyle{plain}{

\fancyhead[C]{\includegraphics[width=18.5cm]{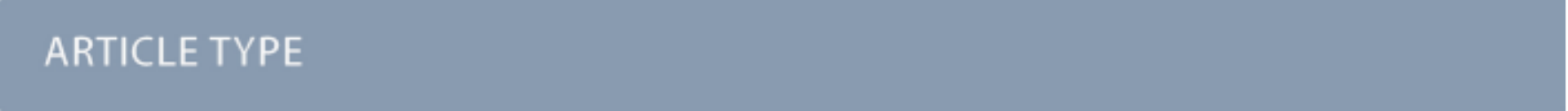}}
\fancyhead[L]{\hspace{0cm}\vspace{1.5cm}}
\fancyhead[R]{\hspace{0cm}\vspace{1.7cm}\includegraphics[height=55pt]{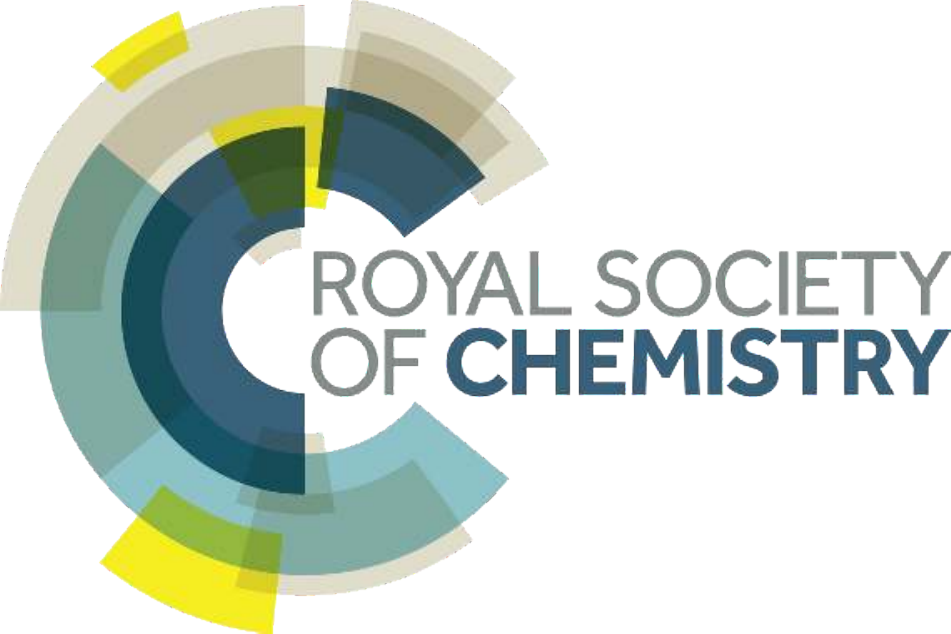}}
\renewcommand{\headrulewidth}{0pt}
}

\makeFNbottom
\makeatletter
\renewcommand\LARGE{\@setfontsize\LARGE{15pt}{17}}
\renewcommand\Large{\@setfontsize\Large{12pt}{14}}
\renewcommand\large{\@setfontsize\large{10pt}{12}}
\renewcommand\footnotesize{\@setfontsize\footnotesize{7pt}{10}}
\makeatother

\renewcommand{\thefootnote}{\fnsymbol{footnote}}
\renewcommand\footnoterule{\vspace*{1pt}%
\color{cream}\hrule width 3.5in height 0.4pt \color{black}\vspace*{5pt}}
\setcounter{secnumdepth}{5}

\makeatletter
\renewcommand\@biblabel[1]{#1}
\renewcommand\@makefntext[1]%
{\noindent\makebox[0pt][r]{\@thefnmark\,}#1}
\makeatother
\renewcommand{\figurename}{\small{Fig.~}}
\sectionfont{\sffamily\Large}
\subsectionfont{\normalsize}
\subsubsectionfont{\bf}
\setstretch{1.125} 
\setlength{\skip\footins}{0.8cm}
\setlength{\footnotesep}{0.25cm}
\setlength{\jot}{10pt}
\titlespacing*{\section}{0pt}{4pt}{4pt}
\titlespacing*{\subsection}{0pt}{15pt}{1pt}

\fancyfoot{}
\fancyfoot[CO]{}
\fancyfoot[RO]{\footnotesize{\sffamily{\thepage}}}
\fancyfoot[LE]{\footnotesize{\sffamily{\thepage}}}
\fancyhead{}
\renewcommand{\headrulewidth}{0pt} 
\renewcommand{\footrulewidth}{0pt}
\setlength{\arrayrulewidth}{1pt}
\setlength{\columnsep}{6.5mm}
\setlength\bibsep{1pt}

\makeatletter
\newlength{\figrulesep}
\setlength{\figrulesep}{0.5\textfloatsep}

\newcommand{\topfigrule}{\vspace*{-1pt}%
\noindent{\color{cream}\rule[-\figrulesep]{\columnwidth}{1.5pt}} }

\newcommand{\botfigrule}{\vspace*{-2pt}%
\noindent{\color{cream}\rule[\figrulesep]{\columnwidth}{1.5pt}} }

\newcommand{\dblfigrule}{\vspace*{-1pt}%
\noindent{\color{cream}\rule[-\figrulesep]{\textwidth}{1.5pt}} }

\makeatother

\twocolumn[
  \begin{@twocolumnfalse}
\vspace{3cm}
\sffamily
\begin{tabular}{m{3.25cm} p{13.0cm} }

 & \noindent\LARGE{\textbf{DNA brick self-assembly with an off-lattice potential}} \\
\vspace{0.3cm} & \vspace{0.3cm} \\
 & \noindent\large{Aleks Reinhardt$^{a}$ and Daan Frenkel$^{a}$} \\

 & \noindent\normalsize{We report Monte Carlo simulations of a simple off-lattice patchy-particle model for DNA `bricks'. We relate the parameters that characterise this model with the binding free energy of pairs of single-stranded DNA molecules. We verify that an off-lattice potential parameterised in this way reproduces much of the behaviour seen with a simpler lattice model we introduced previously, although the relaxation of the geometric constraints leads to a more error-prone self-assembly pathway. We investigate the self-assembly process as a function of the strength of the non-specific interactions. We show that our off-lattice model for DNA bricks results in robust self-assembly into a variety of target structures.} \\

\end{tabular}

 \end{@twocolumnfalse} \vspace{0.6cm}

  ]

\renewcommand*\rmdefault{bch}\normalfont\upshape
\rmfamily
\section*{}
\vspace{-1cm}

\footnotetext{\textit{$^{a}$~Department of Chemistry, University of Cambridge, Lensfield Road, Cambridge, CB2~1EW, United Kingdom. E-mail: ar732@cam.ac.uk, df246@cam.ac.uk.}}


\section{Introduction}

Self-assembling materials have been the subject of considerable scrutiny by researchers.\cite{Halverson2013, Whitelam2015, Frenkel2015} However, most self-assembling structures investigated thus far have been constructed using only a small number of distinct building blocks. The reason is that a system consisting of many different components usually fails to self-assemble due to self-poisoning. It was therefore rather surprising to the community when Peng Yin's group demonstrated that potentially thousands of distinct DNA molecules can reproducibly self-assemble into complex, fully addressable, nearly error-free target structures.\cite{Ke2012, Wei2012} The prospect of such addressable complex self-assembly has captured the imagination of several groups, and much experimental and theoretical work has been undertaken to try and understand the principles and behaviours of such systems.\cite{Hedges2014, Zenk2014, Zeravcic2014, Murugan2015, Whitelam2015b, Madge2015}

In the canonical DNA brick set-up, short single-stranded DNA molecules have sequences chosen in such a way that molecules with which they are designed to hybridise in the target structure are made to be complementary to each other. DNA molecules can hybridise whether or not they are completely complementary; however, the free energy of hybridisation depends strongly on the sequence and Watson--Crick pairing is much more favourable than other combinations of bases. Thus it is generally the case that `designed' interactions (i.e.~those interactions corresponding to hybridisation pairs that are present in the target structure, and which are designed to be complementary) are considerably stronger than all other (`incidental') interactions. This permits the large number of distinct DNA molecules to self-assemble into structures comprising potentially thousands of molecules,\cite{Ke2012} although there is in principle an upper limit to the target structure size on entropic grounds.\cite{Hedges2014}  To try to explain why self-assembly succeeded for such DNA-based structures whilst it failed for many other, similar -- and often even considerably simpler -- systems, we have recently developed simulation-based\cite{Reinhardt2014, Reinhardt2016} and theoretical approaches\cite{Jacobs2015, Jacobs2015b, Jacobs2015c, Jacobs2016} to studying the problem.

The original experiments on DNA bricks\cite{Ke2012} entailed short, 32-nucleotide single-stranded DNA molecules. Each molecule was divided into four domains, with each of the four domains designed to hybridise with a different neighbouring DNA molecule in the target structure. By designing which molecules hybridise with which other molecules, intricate target structures can be designed in a modular manner.\cite{Ke2012} However, the choice of the length of the DNA molecules was not arbitrary: as domains form a double-stranded helix with a length of 8 base pairs, this results in a dihedral angle very close to \ang{90},\cite{Ke2012} since the normal (`B') form of DNA comprises a helix with 10.5 base pairs per turn. This creates a rectangular pattern of DNA helices, but the centres of mass of each of the single-stranded DNA molecules form a (distorted) diamond lattice.\cite{Ke2012} We have previously used this fact to design a very simple `patchy particle' potential, where each particle has four rigid tetrahedrally arranged patches, each with a distinct DNA sequence, to represent the four domains of a DNA molecule.

Our previous simulations with this simple lattice potential have confirmed the experimental hypothesis that nucleation plays a crucial role in the self-assembly process. The underlying idea is simple: at high temperatures, a dilute solution (effectively a `vapour' phase) is thermodynamically stable, whilst at low temperatures, any incidental, undesigned interaction is favoured and large aggregates form instead of the target structure. At intermediate temperatures, incidental interactions are not yet dominant, but designed interactions are sufficiently favourable that the target structure can form. However, there is a free-energy barrier that prevents the target structure from forming \textit{en masse}: the process is initiated by nucleation, which is crucial for the self-assembly process. Since the nucleation of a cluster involves crossing a not insignificant free-energy barrier, it is often described as a rare event. Crucially, this means that the clusters that become post-critical are on average very far apart from each other, meaning that they do not interact and do not have the opportunity to form larger incorrect structures. Moreover, the monomers do not all suddenly rush to form clusters, meaning that the monomers are not depleted too rapidly from the surrounding solution.\cite{Reinhardt2014, Jacobs2015b}

Intriguingly, although this process is not entirely unlike crystal growth, and crystals are well known to form by nucleation, the nucleation behaviour seen in DNA brick self-assembly is non-classical. Unlike in classical nucleation theory, where the nucleating cluster grows without limit once the maximum in the free-energy barrier has been crossed, in the self-assembly of finite structures, the fully assembled structure does not normally appear to be stable at the point at which the nucleation barrier just becomes surmountable.\cite{Jacobs2015b} The reason for this is that whilst there is only a single way to arrange all the particles in the target structure, there are numerous ways of constructing slightly smaller structures, since there are many distinct monomers which can be missing to arrive at a structure of a given incomplete size. At sufficiently high temperatures, this additional entropy wins over the enthalpic favourability of forming the target structure in full. DNA brick structures must therefore be prepared by following a cooling protocol:\cite{Jacobs2015b} once nucleation has occurred, the structure must be cooled further still in order to assemble the target structure to completion.

We have gained a very considerable degree of insight by performing both lattice simulations and theoretical calculations.  However, whilst we have begun to understand the underlying physics which permits DNA brick structures to form, there are several questions that remain unanswered. One particular weakness of the model we have previously proposed is the fact that we have assumed DNA molecules can only move on a lattice and can only adopt one of 24 fixed orientations. A similar constraint was applied in the theoretical approach. Clearly, such constraints have a significant effect on the entropy of the system, and it is therefore important to determine whether the self-assembly that is observed in our lattice model is robust when we go off-lattice. It is not at all clear \textit{a priori} that just because a lattice model forms a finite ordered structure, an off-lattice analogue will as well: a lattice model cannot distinguish between a dense phase that is liquid-like and one that is crystalline. However, the difference between a truly ordered and simply a `dense' structure is crucial in the study of self-assembly.

In this work, we propose a simple off-lattice potential that, while still very much a coarse-grained representation of DNA bricks, can capture more of the translational and orientational entropy of the structural building blocks. We present a general statistical mechanical derivation that results in a simple, yet realistic mapping of the model's parameters to experimental data. Finally, we show that such an off-lattice potential behaves in a way that is analogous to the lattice potential  we had introduced previously and permits us to construct a variety of target structures in a similar manner to that already investigated, but with a few significant differences which we address below.

\section{Matching an off-lattice potential to experimental data}
One of the simplest possible off-lattice potentials that we can use to model the `patchy' nature of the interaction of the DNA bricks introduced above is a Kern--Frenkel-type\cite{Kern2003} potential,
\begin{equation} U(\mathbold{r}_i,\,\mathbold{r}_j,\,\mathbold{\omega}_i,\,\mathbold{\omega}_j) =
\begin{cases}
   \infty & \text{ if } r_{ij} < \sigma,\\
   f(\mathbold{r}_{ij},\,\mathbold{\omega}_i,\,\mathbold{\omega}_j) & \text{ if } \sigma \le
   r_{ij} \le \lambda \sigma,\\
   0 & \text{ otherwise},
\end{cases}
\end{equation}
where $\mathbold{r}_{ij}$ is the interparticle vector of length $r_{ij}$, $\mathbold{r}_i$ and $\mathbold{r}_j$ are the position vectors of particles $i$ and $j$, respectively, and $\mathbold{\omega}_i$ and $\mathbold{\omega}_j$ are their orientations. This is effectively a square well potential, but with an additional angular dependence given by
\begin{equation} f(\mathbold{r}_{ij},\,\mathbold{\omega}_i,\,\mathbold{\omega}_j) =
\begin{cases}
   -\varepsilon & \text{ if } (\hat{\mathbold{r}}_{ij} \cdot \hat{\mathbold{r}}_{i\alpha} \ge \cos \theta_\text{c}) \\ & \quad {} \wedge (\hat{\mathbold{r}}_{ji} \cdot \hat{\mathbold{r}}_{j\beta} \ge \cos \theta_\text{c})  ,\\
   p & \text{ otherwise},
\end{cases}\label{eqn-potential-angularDep}
\end{equation}%
\begin{figure}[tbp]
\centering
\includegraphics{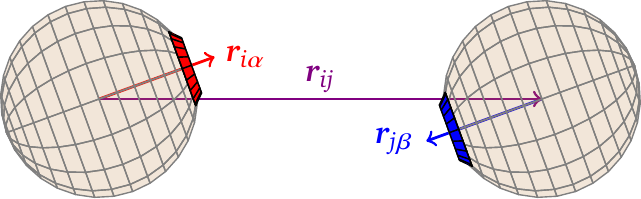}
\caption{Definitions of vectors used in the potential.}\label{fig-vector-defs}
\end{figure}%
where $\hat{\mathbold{r}}_{i\alpha}$ is the normalised position vector of patch $\alpha$ on particle $i$ (see Fig.~\ref{fig-vector-defs}) and $p$ is an optional penalty that can minimise the chance of generating interpenetrating lattices (in our simulations, $p=0$ or $p/k_\text{B}=\SI{100}{\kelvin}$).\cite{FootNote1} The parameter $\sigma$ is the unit of length, whilst $\lambda$, $\varepsilon$ and $\theta_\text{c}$ are parameters yet to be determined.

It is possible to parameterise a potential by finding a suitable mapping between the potential of interest and either experimental data or another potential for which the parameterisation is known; see e.g.~Ref.~\citenum{Bianchi2011b}. In this work, we consider the hybridisation of two single-stranded DNA molecules \ce{A} and \ce{B} to give a hybridised (double-stranded) molecule \ce{AB},
\begin{equation} \ce{A + B <=> AB} ,\label{SL-equilibrium} \end{equation}
for which the equilibrium constant can be written as
\begin{equation}K = \frac{\ce{[AB]}/[\standardstate]}{(\ce{[A]}/[\standardstate])(\ce{[B]}/[\standardstate])} = \frac{\rho_{\ce{AB}} \rho\std}{\rho_{\ce{A}}\rho_{\ce{B}}} = \exp(-\beta \upDelta G\std),\label{eqn-equilibrium} \end{equation}
where $[\standardstate]=\SI{1}{\mole\per\cubic\deci\metre}$ is the standard state concentration, $\rho\std=[\standardstate]N_{\ce{A}} = \SI{6.022e26}{\per\cubic\metre}$ is the standard number density and $\upDelta G\std$ is the standard Gibbs energy for the transformation given in Eqn~\eqref{SL-equilibrium} where 50\,\% of the monomers have hybridised. This hybridisation free energy can be obtained from the SantaLucia thermodynamic model.\cite{SantaLucia2004,FootNoteSantaLucia}

We can write an equivalent expression to Eqn~\eqref{eqn-equilibrium} for the simple Kern--Frenkel model presented above. Assuming that the solution of monomers and dimers is ideal, which is reasonable provided the concentration of each species is small, we can write the canonical partition function of each species $x$ (where $x$ can be \ce{A}, \ce{B} or \ce{AB}) as
\begin{equation}
 Q_x = \frac{V^{N_x}}{N_x! \Lambda_x^{3N_x}} q_x^{N_x},
\end{equation}
where $V$ is the volume of the container, $N_x$ is the number of particles of species $x$, $\Lambda_x$ is the de Broglie thermal wavelength of species $x$ and $q_x$ is the internal partition function of species $x$. Note that the thermal wavelength of \ce{AB} involves integrals over the momenta of both A and B and thus has the dimensions of area rather than length.  Each of the chemical potentials can straightforwardly be calculated from the canonical partition function,
\begin{equation}
\mu_x^{\vphantom{3}} = -k_{\ce{B}} T \pd{\ln Q_x}{N_x}  = k_{\ce{B}} T \ln(\rho_x^{\vphantom{3}} \Lambda_x^3 / q_x^{\vphantom{3}}).
\end{equation}
At equilibrium, $\mu_{\ce{A}}+\mu_{\ce{B}} = \mu_{\ce{AB}}$, which we can solve as
\begin{equation}
\frac{\rho_{\ce{AB}}^{\vphantom{3}}}{\rho_{\ce{A}}^{\vphantom{3}}\rho_{\ce{B}}^{\vphantom{3}}} = \frac{q_{\ce{AB}}^{\vphantom{3}} \Lambda^3_{\ce{A}}\Lambda^3_{\ce{B}} }{q_{\ce{A}}^{\vphantom{3}} q_{\ce{B}}^{\vphantom{3}} \Lambda^3_{\ce{AB}} }.\label{eq-app-equilibrium}
\end{equation}

We assume that the internal state of the monomeric units that bind is not affected by binding. We express this by setting the internal partition functions of the two monomers equal to unity, $q_{\ce{A}}=q_{\ce{B}}=1$.\cite{FootNotePartition} Moreover, because the \ce{AB} molecule is described classically as a dimer of the \ce{A} and \ce{B} particles, the de Broglie thermal wavelengths cancel out, since the momenta of the two monomeric units in the dimer are uncoupled. The rotational partition function of the dimer is thus subsumed into the translational degrees of freedom of the constituting monomers, given that we integrate over the potential energy over all possible states; we include this contribution in the internal partition function $q_{\ce{AB}}$, which will therefore have dimensions of volume. We show in Appendix~\ref{app-dimer-qAB} that it is given by \begin{equation}  q_{\ce{AB}} = \frac{\uppi}{3} \left(\lambda^3-1\right) \sigma^3 (\cos\theta_\text{c} -1)^2  \e^{\beta\varepsilon}. \end{equation}
The equilibrium condition given by Eqn~\eqref{eq-app-equilibrium} can thus be written as
\begin{equation}
\frac{\rho_{\ce{AB}}}{\rho_{\ce{A}}\rho_{\ce{B}}} = \frac{\uppi}{3} \left(\lambda^3-1\right) \sigma^3 (\cos\theta_\text{c} -1)^2  \e^{\beta\varepsilon}.
\end{equation}
Comparing this equation with Eqn~\eqref{eqn-equilibrium} allows us to write
\begin{equation} \frac{\uppi}{3} \left(\lambda^3-1\right) \sigma^3 (\cos\theta_\text{c} -1)^2  \e^{\beta\varepsilon} \rho\std = \exp(-\beta \upDelta G\std).\label{eq-app-exptMatching}
\end{equation}
Using typical dimensions of a DNA brick,\cite{Ke2012} $\sigma^3 \approx \SI{2.5}{\nano\metre}\times \SI{2.5}{\nano\metre}\times \SI{2.7}{\nano\metre}$, gives $\rho\std \sigma^3 \approx 10.1$, leaving only the parameters $\lambda$, $\theta_\text{c}$ and $\varepsilon$ unaccounted for.

\begin{figure*}[tbp]
\centering
\includegraphics{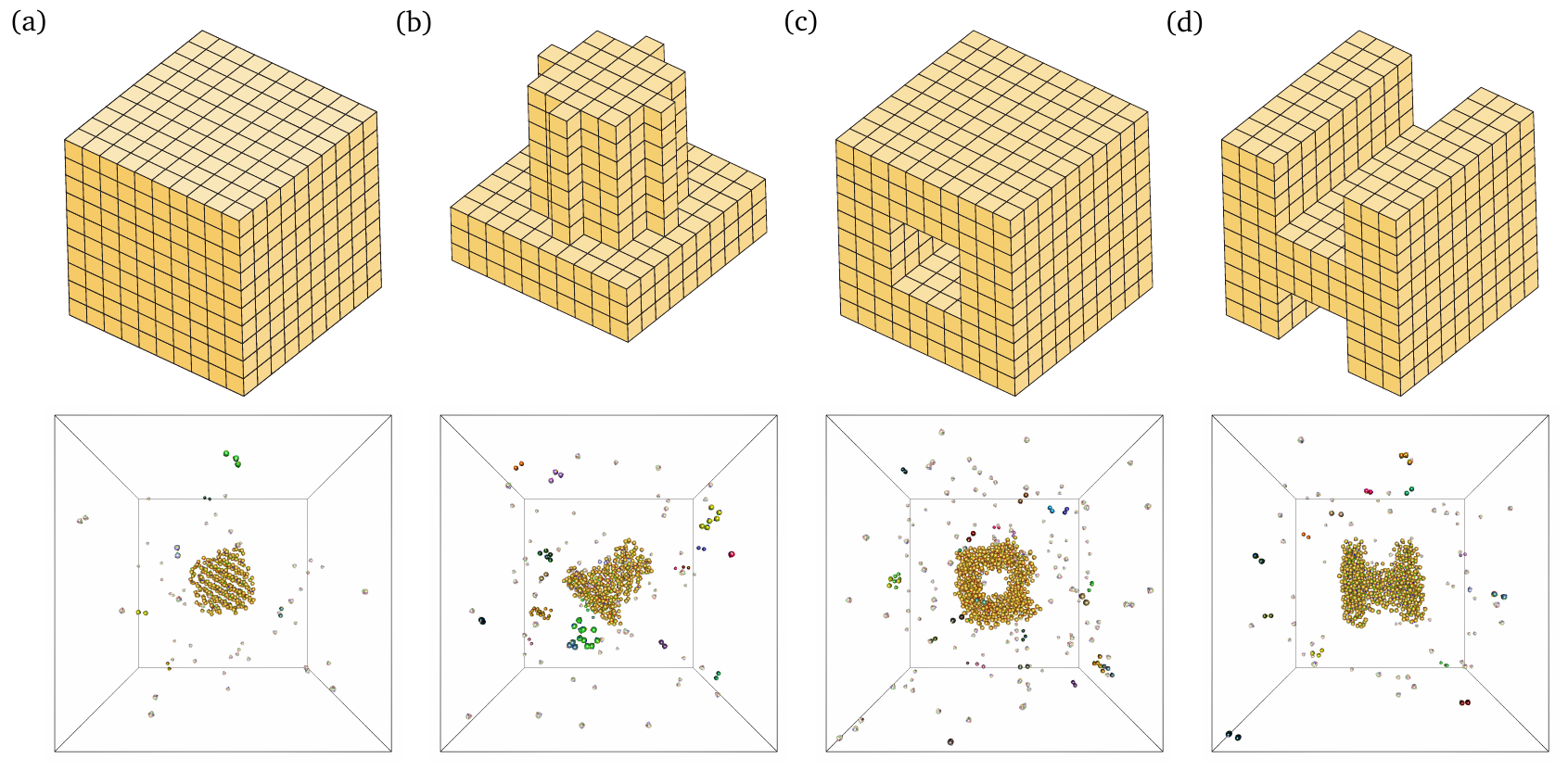}
\caption{Snapshots from brute-force simulations of several structures self-assembled in brute-force Monte Carlo simulations using the off-lattice potential described in the text. A schematic of the designed target structure is also shown for each of the structures. (a) A simple cube. 396 particles in the target structure.  $T=\SI{310}{\kelvin}$,  $\rho\sigma^3=1.48\times10^{-6}$. 
(b) A cylinder on a slab. 489 particles in the target structure. $T=\SI{310}{\kelvin}$, $\rho\sigma^3=1.64\times10^{-6}$. (c) A central cavity structure. 806 particles in the target structure. $T=\SI{314}{\kelvin}$, $\rho\sigma^3=1.64\times10^{-6}$. (d) An H-shaped structure. 696 particles in the target structure. $T=\SI{313}{\kelvin}$, $\rho\sigma^3=1.64\times10^{-6}$. 
}\label{fig-offlattice-structures}
\end{figure*}

Ideally, we might wish to choose $\varepsilon = -\upDelta G\std$. However, at a reasonable bonding distance of $\lambda\sigma=2^{1/3}\sigma=1.26\sigma$, Eqn~\eqref{eq-app-exptMatching} would lead to a patch width of $\theta_\text{c}=\ang{46}$. This is a very wide patch width, which would allow more than one simultaneous patch-patch interaction for any given patch, and thus lead to rather ill-defined structures. Instead, rather than fix $\varepsilon$ and $\lambda$, we can set $\lambda$ and $\theta_\text{c}$ to reasonable values, for example $\theta_\text{c}=\ang{20}$ and $\lambda=1.5$.\cite{FootNote3} We then find that $\varepsilon = -\upDelta G + 2.387\, k_{\ce{B}} T$. In other words, the energy of interaction now accounts for the fact that some entropy is being lost by constraining the bond angle.

The approach we have followed allows us to parameterise an off-lattice potential in a way that captures much of the fundamental physics of the system of interest without introducing a significant bias beyond that of the choice of the form of the potential. However, it ought to be borne in mind that the parameters are not uniquely determined by this mapping. In particular, $\lambda$, $\varepsilon$ and $\theta_\text{c}$ are interdependent. An unreasonably large choice of $\lambda$ or $\theta_\text{c}$ can mean that the assumptions we have made in the derivation can be inappropriate: for example, if more than one particle can bond to a single patch, the dimer assumption is clearly broken. By contrast, a very small patch width or cutoff radius can lead to exceedingly slow dynamics, and so the equilibrium situation may never be reached in simulations. The parameters must therefore be chosen with some consideration given to the practicalities of the required simulations.

\section{Results}

To verify that the model introduced above and parameterised to correspond roughly to experimentally-derived data represents a reasonable approach to simulating DNA brick self-assembly, we perform canonical Metropolis Monte Carlo\cite{Metropolis1953} simulations with `virtual moves'\cite{Whitelam2007, *Whitelam2008} accounting for the motion of clusters. Following the approach we have used with lattice simulations,\cite{Reinhardt2014} we have used umbrella sampling with adaptive weights,\cite{Torrie1977, *Mezei1987} with umbrella sampling steps performed every \num{200000} Monte Carlo steps,\cite{Hetenyi2002} in order to determine the free-energy barrier as a function of the size of the crystalline cluster in the system. Each particle type in the system has four patches arranged in a tetrahedral manner; each patch is assigned a random DNA sequence, but such that patches that point at each other in the target structure have complementary sequences. In every simulation reported here, a single instance of each particle type was placed in the simulation box, so that at most a single copy of the target structure can assemble.

The behaviour we observe is analogous to that seen in lattice simulations, and this in turn has been shown to correspond remarkably well to experimental results.\cite{Reinhardt2014, Jacobs2015b} For example, we are able to self-assemble a range of relatively complex target structures in brute-force simulations, as shown in Fig.~\ref{fig-offlattice-structures}. The underlying behaviour we have proposed for this process in our previous work\cite{Reinhardt2014, Jacobs2015, Jacobs2015b, Jacobs2015c, Reinhardt2016} is still predominantly unchanged: self-assembly in such systems is possible over a limited range of temperatures because of a free-energy barrier to nucleation that prevents immediate aggregation and monomer depletion. The structures shown in Fig.~\ref{fig-offlattice-structures} correspond to some of the largest structures that spontaneously self-assemble in brute-force simulations; while the majority of the target structure can be seen to have formed in each case, the structures are incomplete: as discussed above, the full target structure can be assembled by lowering the temperature after the nucleation process has taken place.

It is noteworthy that for a relatively short-ranged potential such as the one studied here, previous work suggests that the open diamond-like structure is only stable at low pressures and temperatures.\cite{Noya2010} At the temperatures and densities we considered, the work of Romano~\textit{et al.}\cite{Romano2009, Romano2010} suggests that for tetrahedral patchy particles with identical interactions, at equilibrium the mixture phase-separates into a gas and a diamond cubic crystal. In brute-force simulations of patchy particles where every particle is identical and all bonds equally strong, we find that the resulting phase is typically a vapour in equilibrium with a dense fluid, perhaps indicating that the nucleation barrier to forming a diamond-like phase is significant, as expected for patch widths as large as the one we are considering.\cite{Zhang2005, Romano2011} It appears that the fact that each particle is distinct and can only bond strongly with very specific other particles in the system plays a crucial role in enabling us to form tetrahedral structures even in conditions where single-component patchy particles cannot successfully self-assemble.

\begin{figure}[b!]
\centering
\includegraphics{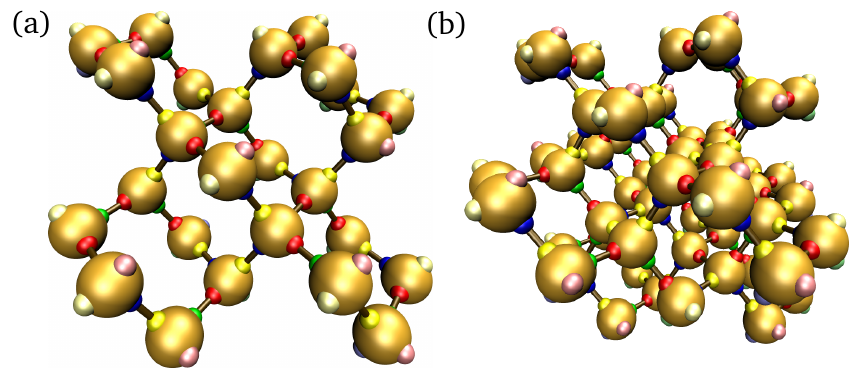}
\caption{Two target structures considered in umbrella sampling simulations. The structure in (a) comprises 26 particles with 32 designed bonds, whilst that in (b) comprises 56 particles with 79 designed bonds. For simplicity, each patch is colour-coded, and by design, red patches bond with yellow ones and blue patches bond with green ones; however, each particle and each patch in the structures are in fact unique. Patches that are bonded in the target structure are shown with a brown `bond'. The outermost patches, shown in paler colours, are passivated by being assigned a poly-T sequence.}\label{fig-offlattice-umbrella-structures}
\end{figure}%
\begin{figure}[t!]
\centering
\includegraphics{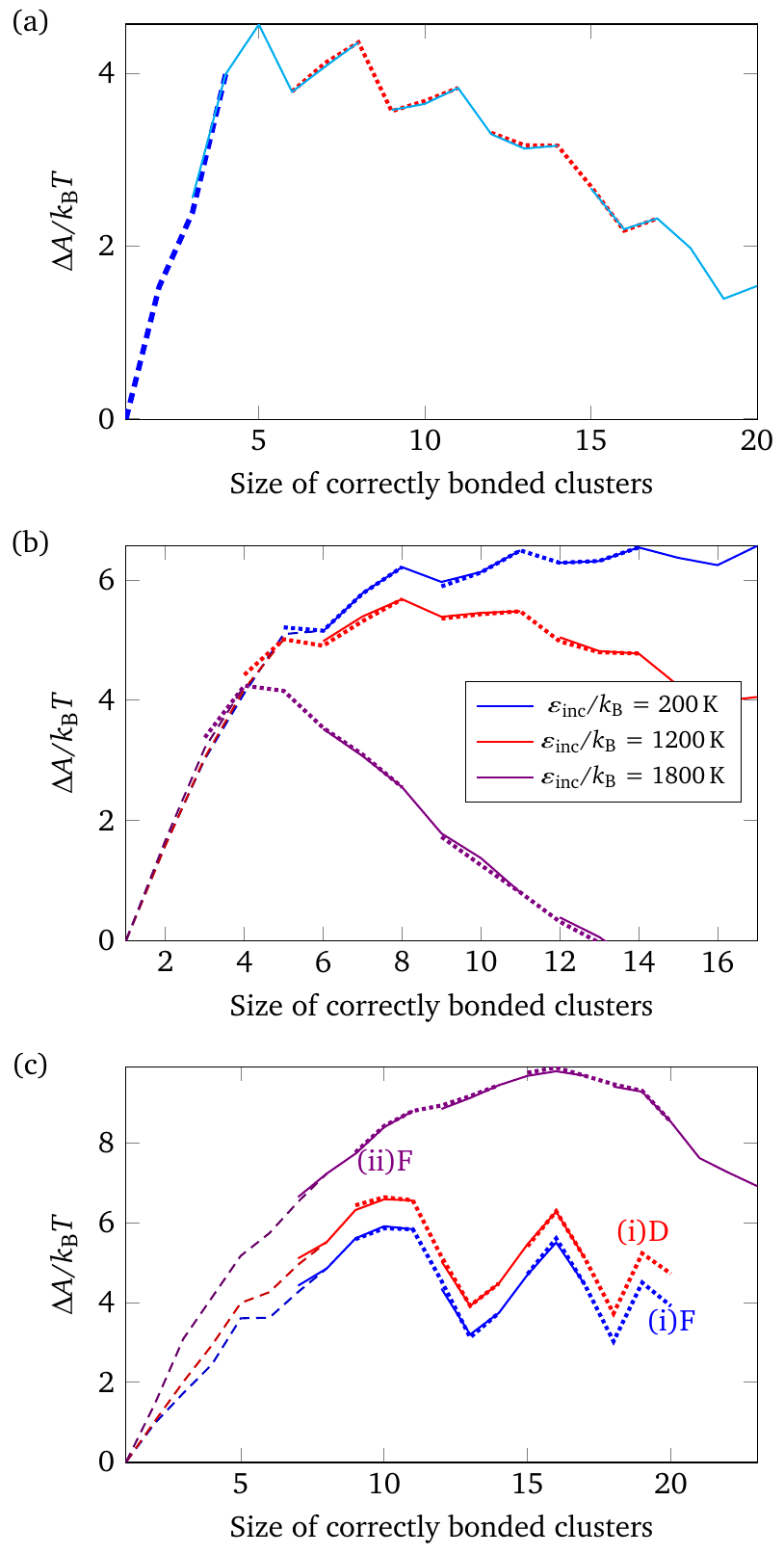}
\caption{The nucleation free energy $\upDelta A$ of the system relative to the vapour of monomers for a very small target structure of 26 particles [Fig.~\refSub{a}{fig-offlattice-umbrella-structures}]. In (a), only designed interactions are included in the energy calculation. $T=\SI{318}{\kelvin}$, $\varepsilon_\text{designed}/k_\text{B} = \SI{4000}{\kelvin}$ (or equivalently $\varepsilon_\text{designed}/k_\text{B}T = 12.58$).  In (b), all designed interactions have a uniform energy of $\varepsilon_\text{designed}/k_\text{B} = \SI{4000}{\kelvin}$ (or equivalently $\varepsilon_\text{designed}/k_\text{B}T = 12.31$), while the incidental interaction strength varies as labelled (in units of $k_\text{B}T$, $\varepsilon_\text{inc}/k_\text{B}T$ is 0.62, 3.69 and 5.54). $T=\SI{325}{\kelvin}$. Alternating line styles are used for the individual umbrella sampling windows and the initial brute force simulation.  In (c), three free-energy profiles corresponding to the full interaction potential, with $\varepsilon$ computed from the SantaLucia model depending on the DNA sequence of each patch, are shown: (i) corresponding to a 26-particle target structure [Fig.~\refSub{a}{fig-offlattice-umbrella-structures}], and (ii) corresponding to a 56-particle target structure [Fig.~\refSub{b}{fig-offlattice-umbrella-structures}]. The label `D' means that only designed interactions were taken into account, whilst `F' indicates that all interactions, designed and undesigned, were taken into account. (i) $T=\SI{308}{\kelvin}$, (ii) $T=\SI{316}{\kelvin}$. $\rho\sigma^3=1.48\times10^{-6}$.}\label{fig-offlattice-freeenergybarrier}
\end{figure}
In addition to brute-force simulations, we have calculated free-energy barriers for small target structures (Fig.~\ref{fig-offlattice-umbrella-structures}) in a range of conditions (Fig.~\ref{fig-offlattice-freeenergybarrier}). It is convenient in the first instance to compute the free-energy barrier for a system in which only the designed interactions are switched on, and they all have the same bonding energy. A free-energy profile for such a system is shown in Fig.~\refSub{a}{fig-offlattice-freeenergybarrier}. The behaviour observed is very similar to that seen in lattice simulations, and the basic features are essentially identical to those observed in lattice simulations\cite{Reinhardt2014} and in theoretical work:\cite{Jacobs2015, Jacobs2015b} the free energy initially increases with the cluster size, as the enthalpic gain of a single bond is insufficient to compensate for the entropic loss of binding a monomer from the vapour phase to the growing cluster. However, the completion of every `cycle', i.e.~a closed loop of particles that are bonded to one another, is a process in which two bonds are formed simultaneously, and this process is thermodynamically favoured. This gives the free-energy barrier as a function of cluster size a distinctive jaggedness, as the free energy decreases upon the formation of individual `cycles' in the largest cluster.

Although this behaviour is expected, the picture changes as interactions between patches that are not bonded in the target structure are switched on. We have investigated this behaviour further by studying a range of systems with pre-determined interaction strengths both for the `designed' and the `incidental' interactions (i.e.~interactions that are present in the target structure and all other possible patch-patch interactions, respectively), whereby all designed patch-patch interactions contribute an energy of $\varepsilon_\text{designed}/k_\text{B} = \SI{4000}{\kelvin}$ and the incidental patch-patch interactions contribute an energy that ranges from $\varepsilon_\text{incidental}/k_\text{B} = \SI{200}{\kelvin}$ to \SI{1800}{\kelvin} (where $\varepsilon$ is defined in Eqn~\eqref{eqn-potential-angularDep}), but in any one simulation all the designed and all the incidental interactions have the same strength.\cite{FootNoteEnergies}  Free-energy profiles for a selection of these systems are shown in Fig.~\refSub{b}{fig-offlattice-freeenergybarrier}. Clearly, the more significant the incidental interactions are, the smoother the free-energy profile becomes. The greater number of possible clusters in off-lattice simulations, including in particular cycles comprising fewer than six monomers, can stabilise the incomplete structures near the top of the free-energy profile in ways that are not possible in on-lattice simulations. Moreover, whilst both the vapour phase and the growing nucleus are stabilised by such incidental interactions, the growing nucleus is stabilised more, reducing the overall height of the nucleation free-energy barrier.\cite{FootNoteReducedHeight}

The free-energy behaviour of systems that can interact via incidental bonds is interesting because it demonstrates that the finer features of the free-energy profile can be lost when studying more realistic systems than the lattice potential we have previously used as a model for DNA brick self-assembly. Furthermore, because the free-energy barrier to nucleation is smaller for off-lattice systems including incidental interactions than it is for on-lattice analogues, the temperature window in which the nucleation barrier is surmountable but incidental interactions are still sufficiently weak for self-assembly to occur is likely to be even smaller than previously estimated. However, the key features of the non-classical nucleation behaviour we have identified previously remain: because the target structure is fully addressable, there is only one possible target structure (even if it may now have more practical realisations because the particles are no longer fixed to lattice sites), whereas there are many possible ways in which to assemble partially formed structures. This means that, in conditions where a free-energy barrier exists to prevent instantaneous nucleation, the target structure is not stable. In order to form the full target structure -- which one can envisage is of crucial importance in experiment, where only the fully formed target structure may exhibit the functionality we desire\mbox{ --,} it is still crucial that a self-assembly protocol be adopted, with the temperature gradually being reduced as the self-assembly proceeds.\cite{Jacobs2015b}

Free-energy barriers for target structures simulated using the full potential described above, with interactions between any two patches, whether `designed' or `incidental', calculated using the longest complementary set of their associated DNA sequences, are shown in Fig.~\refSub{c}{fig-offlattice-freeenergybarrier}. Three free-energy profiles are plotted: the curves labelled (i)F and (i)D correspond to the same choice of DNA sequences, but differ in that the curve labelled D was computed in simulations where only designed interactions were taken into consideration, whilst the curve labelled F corresponds to the full interaction potential, including all incidental interactions. However, the incidental interactions calculated using the DNA sequences associated with each patch are quite weak, and including such weak incidental interactions only slightly stabilises high free-energy structures and thus somewhat reduces the free-energy barrier to nucleation. Finally, the free-energy curve labelled (ii)F in Fig.~\refSub{c}{fig-offlattice-freeenergybarrier} corresponds to a system with a larger target structure. As the target structure size increases, the free-energy barrier to nucleation becomes noticeably smoother, since there are simply many more possible clusters that can form with the same number of building blocks.

The free-energy profiles shown here are not radically different from those we have previously reported for lattice simulations. While the free energy as a function of the largest cluster size is somewhat more difficult to interpret in such off-lattice simulations, it remains the case that the self-assembly is controlled by nucleation, and brute-force simulations confirm that it is still possible to find conditions under which the free-energy barrier to nucleation is sufficiently small that nucleation can occur spontaneously, but large enough to be rate-limiting, as appears to be necessary for successful self-assembly.

\section{Conclusion}
In this work, we have introduced a very simple approach to obtaining a relatively sound parameterisation of a simple off-lattice coarse-grained potential of DNA bricks. In particular, we have shown how a Kern--Frenkel-type potential can be fitted to the hybridisation free energy of two single-stranded DNA molecules that is known from experiment, which allows us to parameterise the potential with comparatively little effort. We have verified that an off-lattice model parameterised in this way gives a reasonable description of the self-assembly of DNA bricks.

The behaviour of DNA bricks that we previously studied using a lattice-based approach both in simulations and using a theoretical approach does not change significantly when simulated using this more realistic off-lattice potential, which helps to support the claim we have previously made that the majority of the underlying physics of self-assembly is captured by the simple patchy model we have previously studied. However, the different dependence on incidental interactions present in the system demonstrates that, not unexpectedly, the off-lattice potential self-assembly is somewhat less robust than its on-lattice analogue. Moreover, comparing the lattice and off-lattice approaches provides us with significant insight into the types of interaction that truly are fundamental and which can safely be coarse-grained away.

Although the computational model we have introduced is still very simple, its off-lattice nature allows us to relax the severe constraints on the geometry of the structures that were able to be assembled using our previous models.\cite{Reinhardt2014, Reinhardt2016} The fact that the underlying self-assembly behaviour is not significantly different when studied using an off-lattice potential is very good news, particularly if a different experimental set-up were to be used to construct the kinds of many-component structures we have investigated. For example, DNA Holliday junctions and multi-arm motifs\cite{Holliday1964, *Lilley2000, *Zhang2008} could be used as building blocks instead of short single-stranded DNA; however, such structures might be expected to be much more floppy than canonical DNA bricks. Although our simulations show that relaxing the geometric constraints that the system must satisfy results in slower, more error-prone assembly, the target structures do in fact form reliably over a narrow range of conditions, which gives us some degree of confidence that alternative experimental strategies are certainly worth exploring in more detail.

In future work, it would be prudent to investigate the initial stages of the nucleation behaviour with a much more realistic model of DNA, perhaps of the order of complexity afforded by oxDNA,\cite{Snodin2015} to verify to what extent the predictions made by our simple coarse-grained potentials are reproduced by DNA. However, this will be a very challenging endeavour indeed, since more realistic potentials are prohibitively expensive to simulate over times sufficiently long to obtain representative behaviour.

\section*{Acknowledgements}
This work was supported by the Engineering and Physical Sciences Research Council [Programme Grant EP/I001352/1]. Research carried out in part at the Center for Functional Nanomaterials, Brookhaven National Laboratory, which is supported by the US Department of Energy, Office of Basic Energy Sciences, under Contract No.~DE-SC0012704. Supporting data are available at the University of Cambridge Data Repository, \href{http://dx.doi.org/10.17863/cam.135}{doi:10.17863/cam.135}.

\appendix

\section{Dimer internal partition function}\label{app-dimer-qAB}
\begin{figure}[tbp]
\centering
\includegraphics{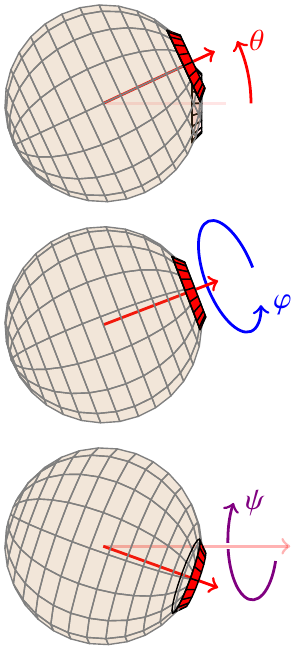}
\caption{Definitions of Euler angles, where the neighbouring particle is implicitly assumed to be placed as in Fig.~\ref{fig-vector-defs}. }\label{fig-angledefs}
\end{figure}

The internal partition function of the dimer corresponds to the volume available for bonding, which is a known result,\cite{Russo2011} but we explicitly derive it here for reference. The internal partition function of the dimer depends on $\mathbold{r}_{\ce{AB}}$, the relative distance of the centres of the monomeric units.  For notational simplicity, we define $\abs{\mathbold{r}_{\ce{AB}}}=r$. We take into account the relative orientations by integrating over both $\mathbold{\omega}_{\ce{A}}$ and $\mathbold{\omega}_{\ce{B}}$. Since the resulting integral is spherically symmetric in $\der \mathbold{r}_{\ce{AB}}$, we can rewrite the volume element in spherical polar co-ordinates as $\der \mathbold{r}_{\ce{AB}} = \der x_{\text{AB}} \, \der y_{\text{AB}}\,\der z_{\text{AB}}= 4\uppi r^2 \,\der r$, giving
\begin{equation} q_{\ce{AB}} = 4\uppi \int r^2\, \der r\, \int \der \mathbold{\omega}_{\ce{A}} \, \int \der \mathbold{\omega}_{\ce{B}}\, \e^{-\beta U}.\label{eqn-integral-over-omega} \end{equation}
To evaluate the remaining integrals over the orientational degrees of freedom, we define the Euler angles $\theta$, $\varphi$ and $\psi$ as shown in Fig.~\ref{fig-angledefs}. The angle $\theta$ measures deviations of the patch position from the interparticle vector; the angle $\varphi$ measures the rotation about the patch position vector; and the angle $\psi$ is the rotation of the patch vector around the interparticle vector. The ranges are thus $\theta\in[0,\,\uppi]$, $\varphi\in[0,\,2\uppi]$ and $\psi\in[0,\,2\uppi]$.
The normalised volume element (Haar measure) for integrating over $\mathbold{\omega}$ is\cite{Vega2008, KosmannSchwarzbach2010}
\begin{equation} \der \mathbold{\omega} = \frac{1}{8\uppi^2} \sin\theta\,\der \theta\,\der \varphi\,\der\psi. \end{equation}
The dot product $\hat{\mathbold{r}}_{\text{A}1}\cdot \hat{\mathbold{r}}_{\text{AB}}$,  which is used in the angular dependence of the Kern--Frenkel potential (Eqn~\eqref{eqn-potential-angularDep}),  is unaffected by rotations about either $\varphi$ or $\psi$; this is to say that the projection of the patch vector onto the interparticle vector remains unchanged by either rotation. The potential energy in the Boltzmann exponent of Eqn~\eqref{eqn-integral-over-omega} thus does not depend on either of these two angles, and we can therefore integrate them out; the orientational volume element is thus given by
\begin{equation} \der \mathbold{\omega} = \frac{1}{2} \sin\theta\,\der \theta. \end{equation}
There are three possible scenarios to consider. Firstly, when $r<\sigma$, $\e^{-\infty}=0$, so there is no contribution to the integral. Secondly, when $r>\lambda \sigma$, the dimer has dissociated, and so this also does not contribute to the internal partition function. The $r$-component of the surviving part of the integral thus satisfies $\sigma\le r \le \lambda \sigma$. In this range, the potential evaluates simply to $-\varepsilon$,
\begin{equation} q_{\ce{AB}} = \frac{4\uppi}{4} \int_{\sigma}^{\lambda\sigma} r^2\,\der r \left[ \int_0^{\theta_\text{c}}  \sin\theta\,\der\theta \right]^2 \e^{\beta\varepsilon}.\label{eqn-integrals-qAB}  \end{equation}
The upper limit for $\theta$ is $\theta_\text{c}$, the patch width; when $\theta>\theta_\text{c}$, the particles no longer form a dimer, so we need not consider that situation.\cite{FootNote2} The remaining integrals in Eqn~\eqref{eqn-integrals-qAB} are readily evaluated, giving
\begin{equation}  q_{\ce{AB}} = \frac{\uppi}{3} \left(\lambda^3-1\right) \sigma^3 (\cos\theta_\text{c} -1)^2  \e^{\beta\varepsilon}, \end{equation}
as used in the main text.


\begin{mcitethebibliography}{47}
\bibitem[FootNoteEnergies()]{FootNoteEnergies}
At temperatures at which nucleation typically occurs for DNA bricks
  ($\sim$\SI{325}{\kelvin}), `designed' interactions correspond to roughly
  $12\,k_\text{B}T$ (i.e.~$\varepsilon_\text{designed}/k_\text{B}\approx
  \SI{3900}{\kelvin}$), whilst incidental interactions correspond to roughly
  $2\,k_\text{B}T$ (i.e.~$\varepsilon_\text{inc}/k_\text{B}\approx
  \SI{650}{\kelvin}$).\relax
\mciteBstWouldAddEndPunctfalse
\mciteSetBstMidEndSepPunct{\mcitedefaultmidpunct}
{}{\mcitedefaultseppunct}\relax
\EndOfBibitem
\bibitem[FootNoteReducedHeight()]{FootNoteReducedHeight}
Reducing the height of the free-energy barrier to nucleation in this way
  generally leads to more errors in self-assembly. The larger the incidental
  interactions, the more defects there are likely to be in the final structure.
  Whether such incidental interactions and the consequent defects are
  acceptable depends on the intended application. For example, assembling an
  approximate target shape is likely to be possible even when there are a
  number of defects in the structure; by contrast, if we were constructing a
  nano-scale circuit board, the addressable nature of the structure would
  almost certainly be extremely important and any defects at all could be very
  problematic.\relax
\mciteBstWouldAddEndPunctfalse
\mciteSetBstMidEndSepPunct{\mcitedefaultmidpunct}
{}{\mcitedefaultseppunct}\relax
\EndOfBibitem
\bibitem[FootNote2()]{FootNote2}
In the simulations reported here, there is a very slight repulsive energy of $p/k_\text{B}=\SI{100}{\kelvin}$ present in our model in order to prevent multiple interpenetrating structures from forming. While this small correction is not accounted for in the derivation presented, the simulation results with and without this additional repulsion are essentially identical and this is not therefore an essential feature of the potential.\relax
\mciteBstWouldAddEndPunctfalse
\mciteSetBstMidEndSepPunct{\mcitedefaultmidpunct}
{}{\mcitedefaultseppunct}\relax
\EndOfBibitem
\end{mcitethebibliography}

\providecommand*{\mcitethebibliography}{\thebibliography}
\csname @ifundefined\endcsname{endmcitethebibliography}
{\let\endmcitethebibliography\endthebibliography}{}

\end{document}